\documentclass[aps,prd, preprint, onecolumn,superscriptaddress,amssymb,amsmath,nofootinbib]{revtex4-1}
\pdfoutput=1

\usepackage{color}
\usepackage{epsfig}
\usepackage{graphicx}
\usepackage{float}
\usepackage{epstopdf}
\usepackage{hyperref} 
\usepackage[usenames,dvipsnames]{xcolor}
\usepackage{slashed}
\usepackage{braket}

\begin{document}

\preprint{EFI-20-2}

\title{A Light Scalar Explanation of $(g-2)_{\mu}$ and the KOTO Anomaly}

\author{Jia Liu}
\affiliation{Physics Department and Enrico Fermi Institute, University of Chicago, Chicago, IL 60637}

\author{Navin McGinnis}
\affiliation{Physics Department, Indiana University, Bloomington, IN 47405, USA}
\affiliation{High Energy Physics Division, Argonne National Laboratory, Argonne, IL, 60439}

\author{Carlos E.M. Wagner}
\affiliation{Physics Department and Enrico Fermi Institute, University of Chicago, Chicago, IL 60637}
\affiliation{High Energy Physics Division, Argonne National Laboratory, Argonne, IL, 60439}
\affiliation{Kavli Institute for Cosmological Physics, University of Chicago, Chicago, IL, 60637}

\author{Xiao-Ping Wang}
\affiliation{High Energy Physics Division, Argonne National Laboratory, Argonne, IL, 60439}


\date{\today}

\begin{abstract}
The KOTO experiment has recently performed a search for neutral Kaons decaying into neutral pions and a pair of neutrinos.   
Three events were observed in the KOTO signal region, with an expected background of about 0.05.  Since no clear signal of systematic errors have been found, the excess of events in the decay $K_{L}\rightarrow\pi^0\nu\bar{\nu}$ is quite intriguing. 
One possibility to explain this anomaly would be the presence of a scalar $\phi$ with mass of the order of the pion mass 
and inducing decays $K_L \to \pi^0 \phi$ which mimic the observed signal.
A  scalar with mass of the order of the pion mass and a coupling to muons of the order of the Standard Model Higgs coupling could also explain the muon anomalous magnetic moment anomaly $(g-2)_{\mu}$. We built on these facts to show that a light singlet scalar with couplings to the leptons and quarks as the ones induced by mixing with Higgs states in two Higgs doublet models may lead to an explanation of both anomalies.
More specifically, we show that this is the case in the so-called type-X models in which leptons and quarks couple to two 
different Higgs doublets, and for scalar masses that are in the range between 40 and 70 MeV. Due to the relatively large coupling to leptons required to fit $(g-2)_{\mu}$, the scalar lifetime accidentally falls into 
the sub-nanosecond range which is essential to evade the severe proton beam dump experiments and astrophysical constraints, though it becomes sensitive to constraints from electron beam dump experiments. 
The additional phenomenological properties of this model are discussed.
\end{abstract}


\maketitle

\tableofcontents




\section{Introduction}

The Standard Model (SM) provides an excellent description of all experimental phenomena~\cite{Tanabashi:2018oca}. After the discovery of the
Higgs, the basic mechanism for electroweak symmetry breaking has been confirmed. Moreover, the Higgs couplings
to the gauge bosons, as well as to the third generation quarks and leptons are in very good agreement with the 
predictions of the SM~\cite{Khachatryan:2016vau}.  Furthermore, no clear signals of new physics have been observed at the LHC, implying
that the SM contains all basic ingredients to be the proper effective theory at the weak scale and perhaps at
much higher energies. 

The SM, however, does not provide an explanation for the observed Dark Matter density in the Universe. More
importantly the couplings of the Higgs to the first and second generation quarks and leptons is still unknown and 
large deviations with respect to the SM predictions may be present in these sectors (see for example, Refs.~\cite{Belanger:2013xza,Perez:2015lra,Coyle:2019hvs}). 
Indeed, it is a priori unlikely that the scalar sector of the theory reduces to a single Higgs doublet. Additional doublets and 
singlets may be present, some of them light and which have not been detected due to their small couplings to
fermions and bosons of the SM.

One example of such scalars is associated with an explanation of the muon $(g-2)$ anomaly. As has been
stressed in many works \cite{Kinoshita:1990aj, Zhou:2001ew, Barger:2010aj, TuckerSmith:2010ra, 
Chen:2015vqy, Liu:2016qwd, Batell:2016ove, Marciano:2016yhf, Wang:2016ggf, Liu:2018xkx}
a light singlet scalar with a mass of the order of the pion mass and a coupling to
muons of the order of $10^{-3}$ may lead to an explanation of this anomaly.  Due to gauge invariance, however, such a singlet, 
cannot couple directly to the muons and could couple to leptons via effective operators involving $SU(2)$ Higgs
doublets. 

Recently, the KOTO experiment looked for the decay of neutral Kaons into neutral pions and a pair of neutrinos \cite{Ahn:2018mvc, KOTO-2019}. In particular, KOTO is mostly sensitive for the process $K_L \to \pi^0 \nu \bar{\nu}$.
The neutral pions would subsequently decay into photon pairs. KOTO performed
a blind search for such events. 
The SM expected rate is more than two orders of magnitude below the current KOTO sensitivity. 
When the process was unblinded, however, three events remain in the signal region, with an expected background
of about 0.05 \cite{KOTO-2019}.

The appearance of such a signal is particularly surprising due to the existence of the so-called Grossman-Nir
bound \cite{Grossman:1997sk}, which is based on a simple relationship between the decay widths of charged and neutral Kaons
into charged and neutral pions and neutrinos. Using the respective lifetimes of charged and neutral Kaons, one obtains that

 \begin{equation}
 {\rm BR}(K_L \to \pi^0\nu \bar{\nu}) \lesssim 4.3 \ {\rm BR}(K^+ \to \pi^+ \nu \bar{\nu})
 \end{equation}
 
This bound puts strong constraints on any high energy physics explanation of the KOTO anomaly.
However, it is known that a scalar, with mass of about the pion mass and either stable (when the
mass is close to the pion mass) or with a  lifetime lower than about a nanosecond can provide an explanation of the KOTO anomaly without violating the Grossman-Nir bound. 
This is due to the experimental sensitivities of the charged and neutral Kaon experiments. 
The possibility of a stable, light scalar, with mass
close to the pion mass, leading to a possible excess at the KOTO experiment without violating the 
Grossman-Nir bound was first stressed in Refs.~\cite{Fuyuto:2014cya, Hou:2016den}, well before the KOTO excess observation. 
The possibility that a light scalar, with mass different from the pion mass and  with a long lifetime, but not necessarily stable, can also explain the observed KOTO excess 
without violating the Grossman Nir bound was stressed by several authors, including those of Refs.~\cite{Kitahara:2019lws, Egana-Ugrinovic:2019wzj, Dev:2019hho}. Other possibilities have been discussed in Refs.~\cite{Kitahara:2019lws, Fabbrichesi:2019bmo, Li:2019fhz, Jho:2020jsa}.
 
The range of scalar masses necessary to explain the KOTO anomaly  is very similar to the one arising in the light scalar explanation of the $(g-2)_\mu$ anomaly,  suggesting a 
possible common explanation of both physical phenomena.  In this article, we propose a simple model, which leads to such an explanation. Our model is based on a singlet scalar
mixing with Higgs doublet states as the ones that appear in  a type-X two Higgs doublet model (2HDM), in which quarks and leptons couple to different Higgs doublets. The required singlet scalar turns out to be lighter than the muon and pion. 

Our model differs from previous proposals to explain the KOTO anomaly \cite{Egana-Ugrinovic:2019wzj, Dev:2019hho}, in which a singlet scalar mixing with the SM Higgs (SSM) is assumed. The decay of the scalar into electrons in such proposals lead to a lifetime of the order of  $c\tau \sim 100$~km due to the small mixing angle $\sin\theta \sim 10^{-3}$ necessary to explain the KOTO anomaly. In our scenario, instead, the lifetime is much shorter. It is indeed a non-trivial fact that the coupling to leptons consistent with the explanation of $(g-2)_\mu$ leads to the proper scalar lifetime of $\sim 0.01 $ meter that allows to avoid the charged pion decay constraints. Moreover, this fact also allows a solution to the KOTO anomaly in a different mass range compared with previous proposals. 

The article is organized as follows. In section \ref{sec:effmodel}, we introduce the effective model and show how it can explain both the
$(g-2)_\mu$ and KOTO anomalies. In section \ref{sec:UVmodel}, we discuss the ultraviolet (UV) completion of the model and its constraints from
collider, beam dump and astrophysics searches. We reserve section \ref{sec:conclusion} for our conclusions.

\section{The effective model for the $(g-2)_{\mu}$ and KOTO anomalies}
\label{sec:effmodel}

The anomalous magnetic moment of the muon is an excellent probe of new physics at the weak scale. 
It is governed mainly by a dimension five dipole operator, which is chirally violating, whose contribution to $a_\mu =(g-2)_\mu/2 $ is expected to be loop suppressed and proportional to $(m_\mu/\tilde{m})^2$,
where $\tilde{m}$ parameterizes the scale of new physics. The current measurement is more than
three standard deviations from the expected value in the SM~\cite{Blum:2018mom}, namely

\begin{equation}
\label{gm2_dev}
\Delta a_{\mu} \equiv a_{\mu}^{exp} - a_{\mu}^{SM}=(2.74 \pm 0.73) \times 10^{-9}
\end{equation}

If this deviation were confirmed, it would be a clear indication of new physics at or below the weak scale. 
A new measurement of $a_\mu$ is expected to be reported by the $g-2$ collaboration at Fermilab within the next few months \cite{Grange:2015fou}.
Many models of physics beyond the SM can lead to an explanation of this anomaly \cite{Miller:2007kk, Jegerlehner:2009ry}. 
In our work, we consider a simple extension of the SM which includes a light scalar $\phi$ that couples to both the leptons and quarks but with different couplings relative to the SM ones.
We assume that the couplings are flavor diagonal and universal within each sector. Thus, we consider the low-energy effective theory defined by

\begin{align}
	\quad \mathcal{L}_{\rm eff} \supset  \sum_{q} \epsilon_{q} \frac{m_q}{v} \phi \bar{q}q + 
	\sum_{\ell} \epsilon_{\ell} \frac{m_\ell}{v} \phi \bar{\ell} \ell + \epsilon_W  \frac{2 m_W^2}{v}\phi W_\mu^+ W^{\mu -}.
	\label{eq:eff-lag}
\end{align}

This setup can occur, for instance, in a lepton-specific 2HDM with an additional singlet. Later, we will argue that $\epsilon_W \approx \epsilon_q$ 
in some reasonable limit. Thus, comparing with the SSM model, the only difference with our effective model is that $\epsilon_{\ell}$ 
is an extra free parameter. We will return to the UV completion of this model in the next section.
In \cite{Liu:2018xkx} it was shown how a light scalar with a generic coupling to the muon can account for the deviation in the muon anomalous magnetic moment. 
The contribution to $(g-2)_{\mu}$ in our effective model is given by \cite{PhysRevD.5.2396, Leveille:1977rc, Lindner:2016bgg}
\begin{equation}
\Delta a_{\mu} = \frac{m_{\mu}^2}{8\pi^2v^2}\epsilon_{\ell}^2\int_{0}^{1}dx\frac{(1-x)^2(1+x)}{(1-x)^2 + x(m_{\phi}/m_{\mu})^2}.
\end{equation}
It is easy to calculate that the requirement to satisfy $(g-2)_{\mu}$ fixes $\epsilon_{\ell}\simeq\mathcal{O}(1)$, for the range of masses $m_{\phi}$ that is relevant for the KOTO anomaly.

In a recent paper \cite{Kitahara:2019lws}, it was shown that the KOTO anomaly could be explained by new physics involving a particle, $X$, with lifetime $\tau\sim\mathcal{O}(0.1-0.01)$ ns and appearing in decays of Kaons with a neutral Kaon branching ratio $ {\rm BR}(K_L \rightarrow\pi^0 X)\sim 10^{-2}$--$10^{-8}$. In our case, the decay of the Kaon is induced at one-loop through penguin diagrams with the $W$ boson and leads an effective $s-d-\phi$ vertex. The partial decay widths for $K_L$ are then controlled by $\epsilon_q$ and $m_{\phi}$ \cite{Leutwyler:1989xj, Gunion:1989we},

\begin{align}
	\Gamma(K_L \rightarrow \pi^0 \phi)=&\frac{\left(\text{Re} \left[g(\epsilon_{q}) \right] \right)^2}{16\pi m_{K}^3}\lambda^{1/2}(m_{K}^2, m_{\pi}^2, m_{\phi}^2),  \\
	g(\epsilon_{q})=&\frac{3m_{K}^2}{32\pi^2 v^3}\epsilon_{q} f_{+}(0) \sum_{q = u,c,t} m_{q}^2V^{*}_{qd}V_{qs} ,
\end{align}
where $\lambda(x,y,z) = x^2+y^2+z^2 - 2 x y-2 y z - 2 x z$ is the triangle function and $f_{+}(0)$ is the
vector form factor at zero momentum transfer. For the partial width of $K^+ \to \pi^+ \phi$, one needs to substitute $\text{Re} \left[g(\epsilon_{q}) \right]$ to $\left|g(\epsilon_{q}) \right|$ and change the corresponding mass parameters.
The calculation of the decay width involves the evaluation of the appropriate form factor. Following Ref.~\cite{Gunion:1989we} (and references therein) and also validated in Ref.~\cite{Leutwyler:1989xj} 
using Higgs low-energy theorems, this factor is taken to be close to unity.
A more precise calculation from lattice QCD shows $f_{+}(0) = 0.9709(46)$,  which is indeed
close to unity \cite{Carrasco:2016kpy}.

\begin{figure}[htb]
	\includegraphics[scale=0.75]{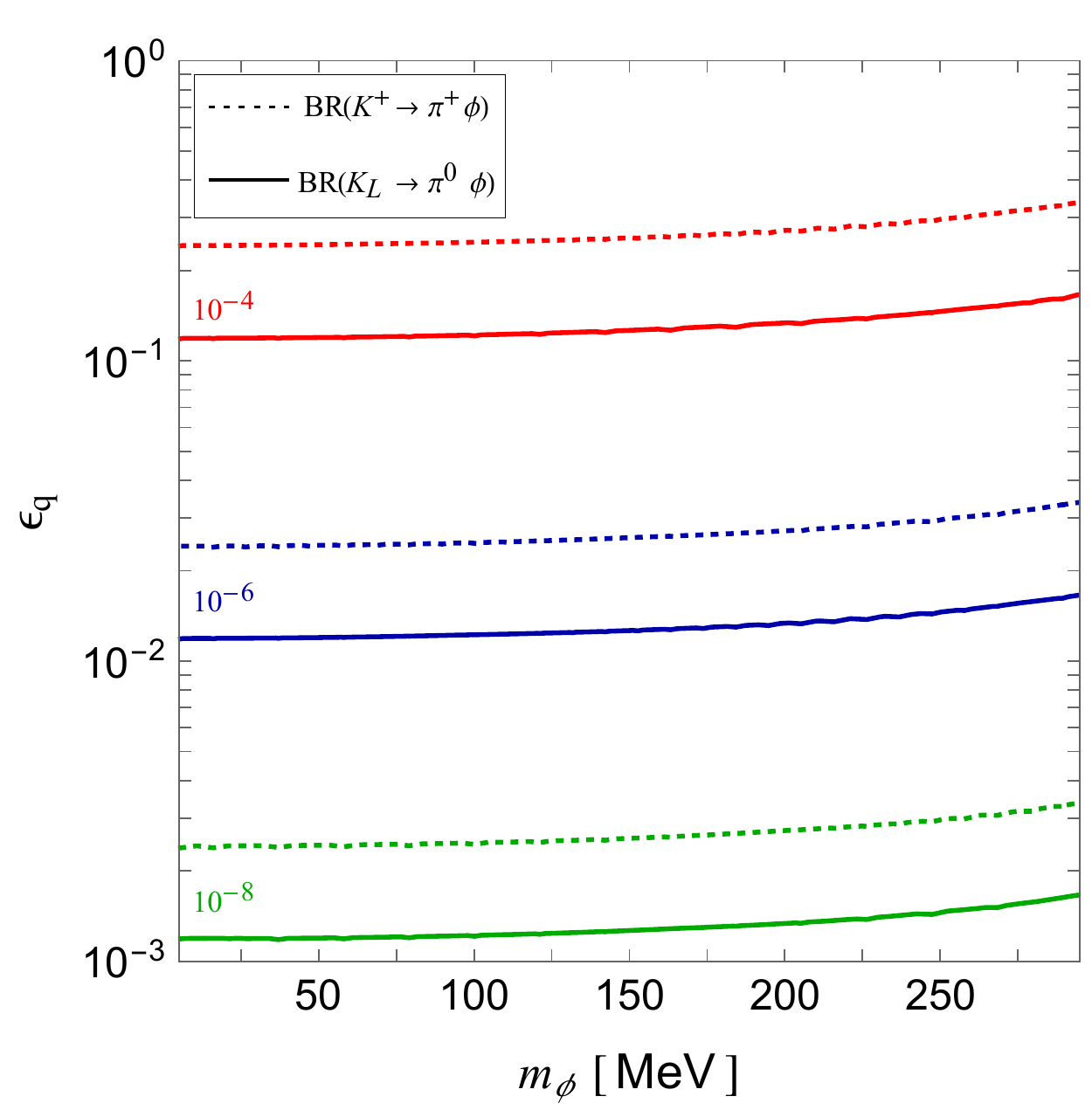}
	\caption{Contours for ${\rm BR}(K_L \rightarrow\pi^0 \phi)$ and ${\rm BR}(K^+ \rightarrow\pi^+ \phi)$ in the $\epsilon_q$--$m_{\phi}$ plane. The solid colored lines indicate the contours for ${\rm BR}(K_L \rightarrow\pi^0 \phi)$. Dotted lines of corresponding colors show where ${\rm BR}(K^+ \rightarrow\pi^+ \phi)$ achieves the corresponding value.}
	\label{fig:kaon-BR}
\end{figure}

In Fig. \ref{fig:kaon-BR}, we show the contours for ${\rm BR}(K_L \rightarrow\pi^0 \phi)$ and ${\rm BR}(K^+ \rightarrow\pi^+ \phi)$ (solid and dotted lines respectively) in the $m_{\phi} - \epsilon_q$ plane.  We see that in the range of scalar masses we consider, the branching ratio is fairly insensitive to $m_{\phi}$ and thus determined mostly by $\epsilon_{q}$. Furthermore, obtaining a branching ratio appropriate for the KOTO anomaly sets $\epsilon_{q} \ll \epsilon_{\ell}$, as the latter is of order 1.

\begin{figure}[htb]
	\includegraphics[scale=0.75]{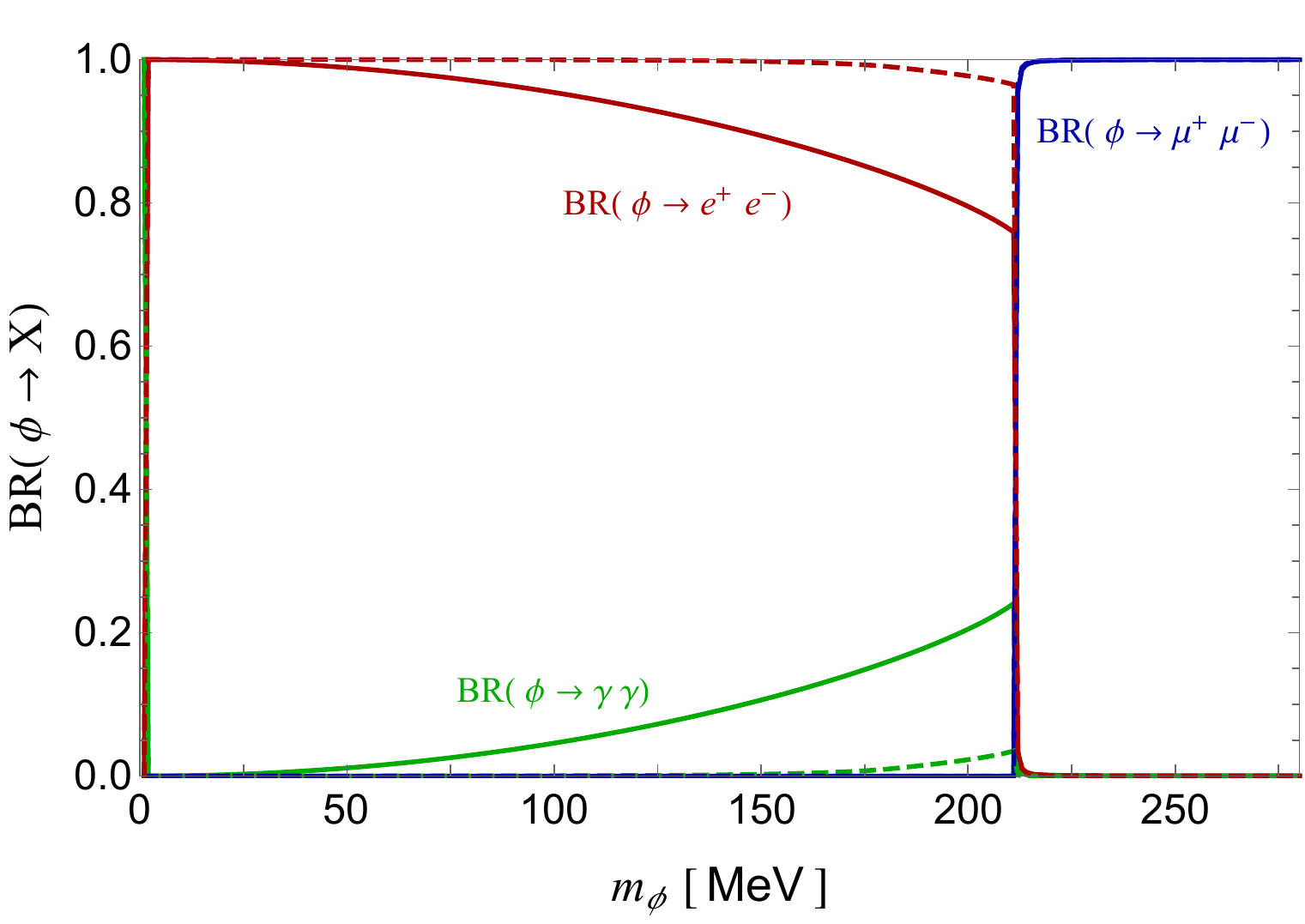}
	\caption{Branching ratios of the allowed $\phi$ decay modes. The solid lines show branching ratios ${\rm BR}(\phi\rightarrow X)$ when  the coupling $\epsilon_{\ell} \sim 1$ is fixed to fit the central value for $\Delta a_{\mu} $, and $\epsilon_{q} = \epsilon_{W} \sim 6\times 10^{-3}$ is fixed so that ${\rm BR}(K_L\rightarrow \pi^0 \phi)=10^{-6}$. In dashed lines, the branching ratios for singlet scalar mixing model $\epsilon_{q}=\epsilon_{W} =\epsilon_{\ell}$ are shown. In the latter case, a dramatic cancellation between $W$ loop function and fermion loop function happens for $m_\phi$ around the pion mass, which leads to a higher electron-positron decay branching ratio even with a much smaller $\epsilon_\ell$. 
	}
	\label{fig:BR_modes}
\end{figure}

With $\epsilon_{\ell}$ fixed by the requirement to explain $(g-2)_{\mu}$ and $\epsilon_{q}$ fixed by the choice of ${\rm BR}(K_L \rightarrow\pi^0 \phi)$ to explain the KOTO experiment result, the lifetime of $\phi$ is then uniquely determined by its mass, once all allowed decay modes are identified. 
For the range of masses relevant to both experiments, the allowed decay modes are $\phi\rightarrow e^{+}e^{-}$, $\phi\rightarrow\gamma\gamma$, and when $m_{\phi}>2m_{\mu}$ $\phi\rightarrow\mu^{+}\mu^{-}$.
The decay widths are given by 
\begin{align}
& \Gamma(\phi\rightarrow \ell\ell)=\frac{\epsilon_{\ell}^2m_{\ell}^2}{8\pi v^2}m_{\phi}(1-\tau_{\ell})^{3/2}\theta({m_{\phi}^2 - 4m_{\ell}^2}),
\label{eq:tree} \\
& \Gamma(\phi\rightarrow\gamma\gamma)=\frac{\alpha^2 m_{\phi}^3}{1024\pi^3}\left| \sum_{q}\frac{6\epsilon_{q}}{v}Q_{q}^2A_{1/2}(\tau_{q})  + \sum_{\ell}\frac{2\epsilon_{\ell}}{v}A_{1/2}(\tau_{\ell}) + \frac{2\epsilon_{W}}{v}A_{1}(\tau_{W})\right|^2,
\label{eq:diphoton}
\end{align}
where  $\theta$ is the step function, $A_{1/2}(\tau_{i})$ ($A_{1}(\tau_{i})$) is the usual fermion (vector-boson) loop function and $\tau_{i}=4m_{i}^2/m_{\phi}^2$ \cite{Gunion:1989we}. 

In Fig. \ref{fig:BR_modes}, we show the branching ratios of the allowed decay modes of $\phi$. The decay widths are determined by fixing $\epsilon_{\ell}$ so that $\Delta a_{\mu}$ is fit to the central value of $(g-2)_\mu$, and $\epsilon_{q}$ so that ${\rm BR}(K_L \rightarrow \pi^0 \phi)=10^{-6}$. However, the branching ratios are largely unaffected by deviations from these choices so long as $\epsilon_{q} \ll \epsilon_{l}$. Despite the fact that the diphoton width receives contributions from both quarks and leptons it is loop suppressed, and thus $\phi\rightarrow e^{+}e^{-}$ will always be the dominant decay mode for $m_{\phi}<2 \ m_{\mu}$. Besides the decay branching ratio of our model (solid lines), we also show the corresponding branching ratios in the SSM model (dashed lines) which has $\epsilon_q = \epsilon_\ell =\epsilon_W$. It is interesting to mention that for our model $\epsilon_q  =\epsilon_W \sim 10^{-3} \textup{--} 10^{-2}$ and $ \epsilon_\ell \sim 1$, thus it is natural to expect ${\rm BR}(\phi \to e^+ e^-) \simeq 1$. However, due to a surprising cancellation between $W$ loop function and fermion loop function contributions to the di-photon decay amplitude
for $m_\phi$ around the pion mass, the SSM model has a larger ${\rm BR}(\phi \to e^+ e^-)$. Therefore, in some beam dump experiments our model will be more constrained.

\begin{figure}[htb]
	\includegraphics[width= 0.8 \textwidth]{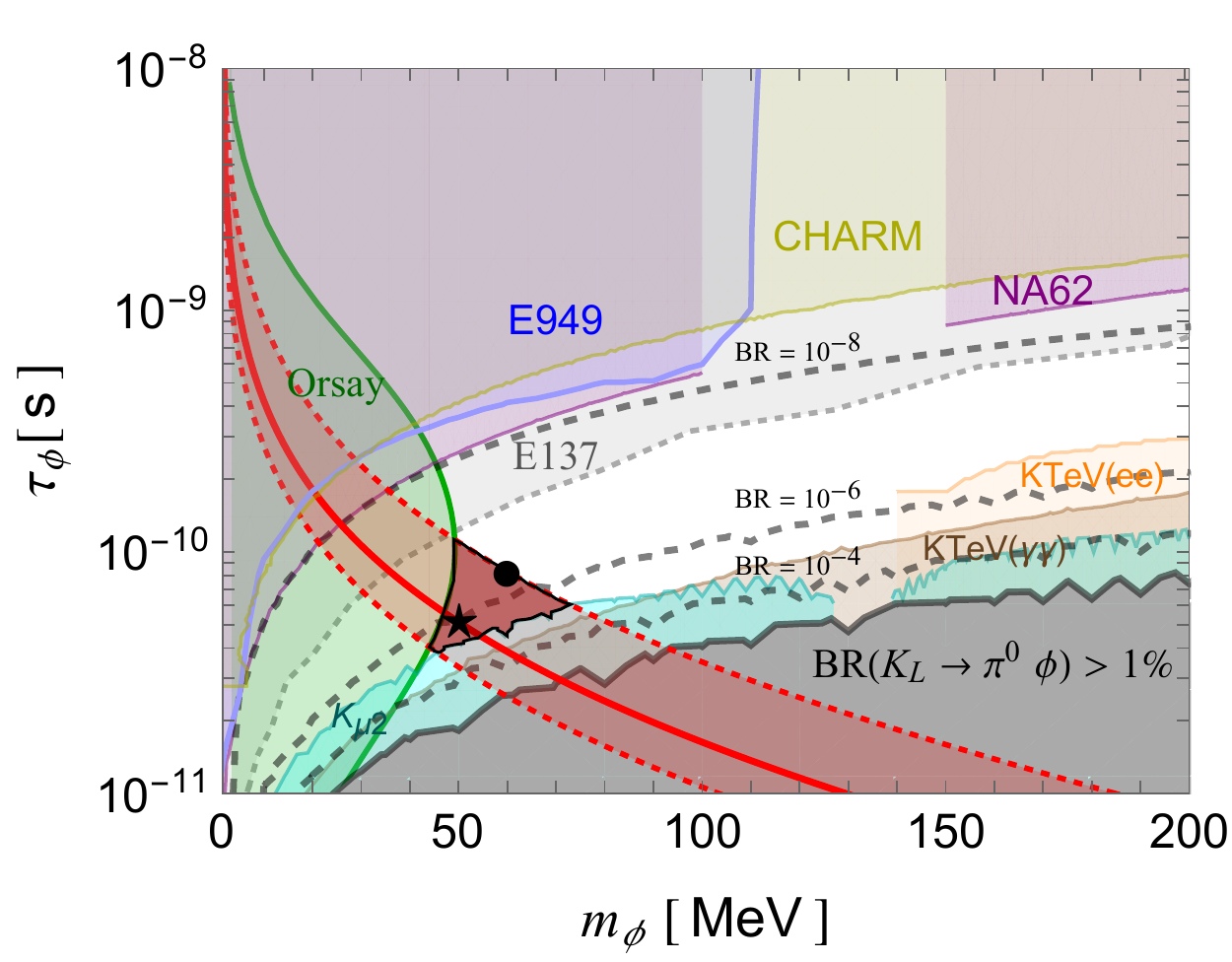}
	\caption{The parameter space for the effective model in the $m_{\phi}$--$\tau_{\phi}$ plane. For each point on the plane, ${\rm BR}(K_L \rightarrow \pi^0 \phi)$ is fixed to fit to the central value measured by the KOTO experiment. We show contours of ${\rm BR}(K_L \rightarrow \pi^0 \phi) = 10^{-4},10^{-6}, \text{ and } 10^{-8}$ in gray dashed lines.
	Once the value of $\epsilon_{\ell}$ is fixed to fit $(g-2)_\mu$, the scalar lifetime shrinks to the red shaded region, where the red solid line corresponds to the central value of $\Delta a_\mu$ and the red dotted lines correspond to the $2 \sigma$ region. 
	The dark red highlighted region can explain both $(g-2)_\mu$ and KOTO anomalies.
	The cyan, blue, purple, yellow, light orange, brown, green, and light gray shaded regions are excluded at $95\%$ C.L. by the $K_{\mu 2}$ \cite{Yamazaki:1984vg}, E949 \cite{Anisimovsky:2004hr, Artamonov:2009sz}, NA62 \cite{NA62-2019}, CHARM \cite{Bergsma:1985qz}, KTeV/E799 ($K_L \to \pi^0 e^+e^-$) \cite{AlaviHarati:2003mr}, KTeV ($K_L \to \pi^0 \gamma\gamma$) \cite{Abouzaid:2008xm}, Orsay \cite{Davier:1989wz}, and E137 \cite{Bjorken:1988as} experiments respectively.
	The gray shaded region is excluded by untagged $K_L$ decay \cite{Tanabashi:2018oca}.
	}
	\label{fig:major-result}
\end{figure}

\begin{figure}[htb]
	\includegraphics[width= 0.48 \textwidth]{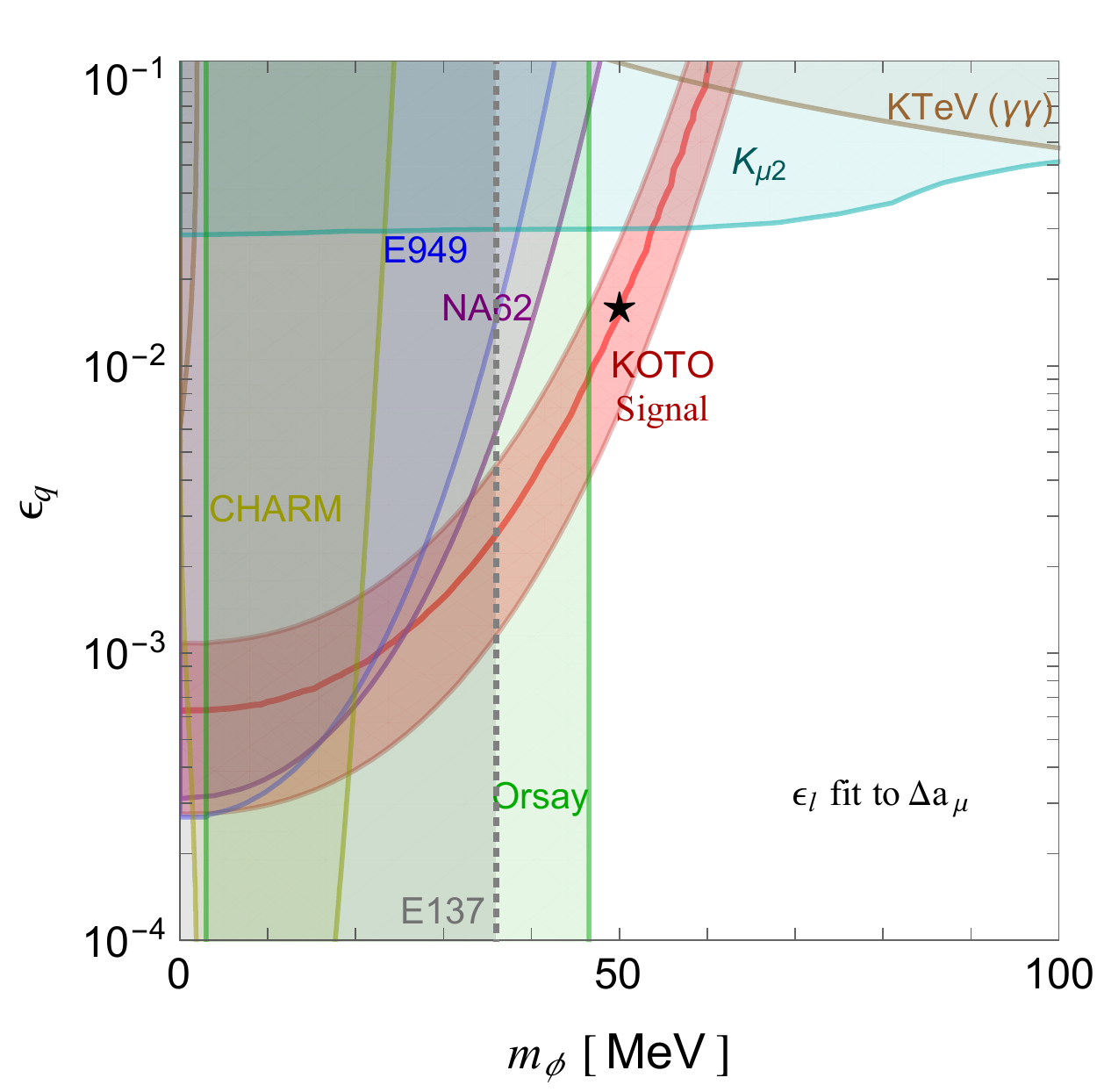}
	\includegraphics[width= 0.48 \textwidth]{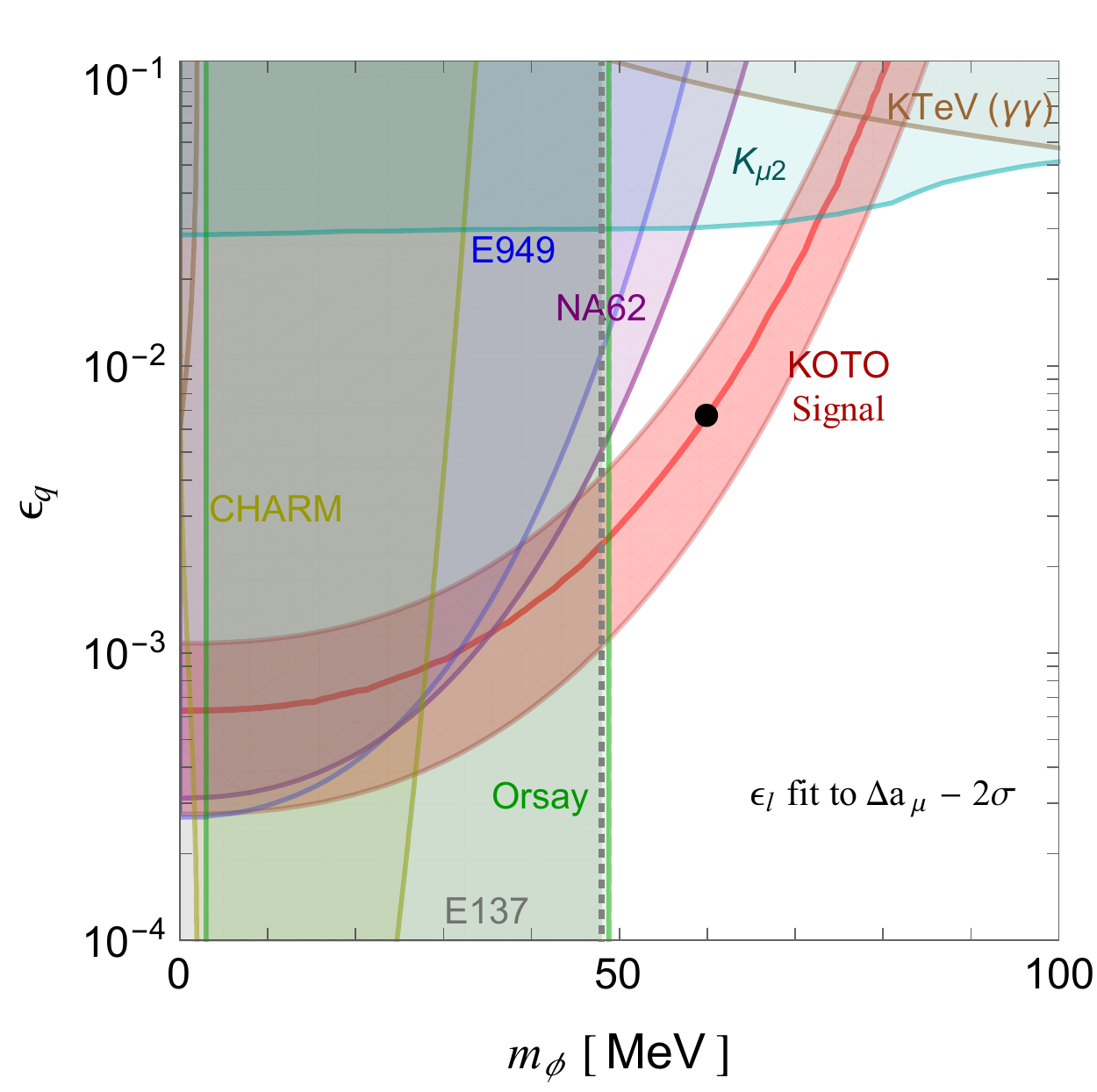}
	\caption{{\bf Left:} The parameter space for the effective model in the $m_{\phi}$--$\epsilon_{q}$ plane when $\epsilon_\ell$ is fixed for the central value of $\Delta a_\mu$. 
	The red shaded band shows the parameter space where the KOTO signal is achieved at $95\%$ confidential level. {\bf Right:} The parameter space for the effective model in the $m_{\phi}$--$\epsilon_{q}$ plane when $\epsilon_\ell$ is fixed to $\Delta a_\mu - 2\sigma$.
	All the constraints are shown in the same colors as in Fig. \ref{fig:major-result}. 
	}
	\label{fig:major-result-2}
\end{figure}

For a light scalar to mimic the signal in the KOTO experiment, it must decay outside of the detector. Taking into account the appropriate boost factors and detector efficiencies this leads to an effective branching ratio
\begin{equation}
{\rm BR}(K_L \rightarrow \pi^0 \phi; {\rm KOTO}) = \epsilon_{{\rm eff}} {\rm BR}(K_L \rightarrow \pi^0 \phi) e^{-\frac{L}{p_\phi}\frac{m_{\phi}}{\tau_{\phi}}},
\label{eq:BR_KOTO}
\end{equation}
where $p_\phi$ is the typical momentum of $\phi$ and $L$ is the detector size in the KOTO experiment. The efficiency factor $\epsilon_{ \rm eff}$ is included to account for the signal efficiency difference between $K_L \rightarrow \pi^0 \phi$ and $K_L \rightarrow \pi^0 \bar{\nu} \nu$, which is taken from \cite{Ahn:2018mvc}. Following \cite{Kitahara:2019lws}, we fix $L=3$m and $E_{\phi}=1.5$ GeV for the KOTO experiment.

In our model, there are three free parameters $m_\phi$, $\epsilon_{\ell}$ and $\epsilon_{q} = \epsilon_{W}$ where the last two are fixed by the $(g-2)_\mu$ and KOTO experiments, respectively. One can trade them into another set of parameters $m_\phi$, $\tau_{\phi}$ and ${\rm BR}(K_L \rightarrow \pi^0 \phi)$. Since $\epsilon_{q} \ll \epsilon_{\ell}$, the total width and lifetime of $\phi$ are dominantly determined by $\epsilon_{\ell}$. In this sense, $\tau_{\phi}$ is insensitive to the precise value of ${\rm BR}(K_L \rightarrow \pi^0 \phi)$. 

In Fig. \ref{fig:major-result}, we show the model parameter space in the $m_\phi$--$\tau_{\phi}$ plane.
Once $\epsilon_{\ell}$ is fixed to fit $(g-2)_\mu$, the model parameters shrink to the red shaded region, where the red solid line corresponds to the central value of $\Delta a_\mu$ and the red dashed lines correspond to the $2 \sigma$ region. 
The dashed gray lines show contours of ${\rm BR}(K_L\rightarrow \pi^0\phi)$ where the central value of the KOTO signal is achieved (see Fig.~\ref{fig:kaon-BR} to translate this to values of $\epsilon_{q}$).
The cyan, blue, purple, yellow, light orange, brown, green, and light gray shaded regions are excluded by the $K_{\mu 2}$ \cite{Yamazaki:1984vg}, E949 \cite{Anisimovsky:2004hr, Artamonov:2009sz}, NA62 \cite{NA62-2019}, CHARM \cite{Bergsma:1985qz}, KTeV/E799 ($K_L \to \pi^0 e^+e^-$) \cite{AlaviHarati:2003mr}, KTeV ($K_L \to \pi^0 \gamma\gamma$) \cite{Abouzaid:2008xm}, Orsay \cite{Davier:1989wz}, and E137 \cite{Bjorken:1988as}  experiments respectively.
The gray shaded region is excluded by untagged $K_L$ decay \cite{Tanabashi:2018oca}. 
We defer the details of the constraints to section \ref{sec:UVmodel}. The parameter space which can solve both the $(g-2)_\mu$ and KOTO anomalies is highlighted in the dark red region near $m_{\phi}\sim 50$ MeV.

In Fig. \ref{fig:major-result-2} we show our results in the $m_\phi$--$\epsilon_{q}$ plane. In the left plot, we show in the red shaded region where the KOTO signal is achieved at the $95\%$ C.L. when $\epsilon_\ell$ is fixed for the central value of $\Delta a_\mu$. When $\epsilon_\ell$ is allowed to vary within a $2\sigma$ range of $\Delta a_\mu$, the variation will be reflected in the lifetime of $\phi$. This will modify the probability decay factor in both the KOTO effective branching ratio and the corresponding factors in the constraints. The right plot of Fig. \ref{fig:major-result-2} shows the result of fixing $\epsilon_\ell$ to $\Delta a_\mu - 2\sigma$, displaying the maximum range allowed for $m_{\phi}$ which satisfies the experimental constraints. \footnote{Similarly, fixing $\epsilon_\ell$ to fit $\Delta a_\mu + 2\sigma$ would give an indication of the minimum $m_{\phi}$ allowed. However, from Fig. \ref{fig:major-result} we see that this range (bottom orange dashed curve) is ruled out by multiple constraints. }

In Fig.~\ref{fig:major-result} and \ref{fig:major-result-2}, we see that the allowed mass range for $\phi$ which can explain both $(g-2)_{\mu}$ and 
KOTO anomalies fall into the range of $\sim 40 - 70$ MeV. In Table~\ref{tab:eff-model_benchmark}, we give two benchmark points for the model parameters. 
Both points are indicated in Figs.~\ref{fig:major-result} and \ref{fig:major-result-2}. 
We have further checked the Barr-Zee diagram contribution from the top quark and the tau lepton for these benchmark points~\mbox{
\cite{Giudice:2012ms, Davoudiasl:2018fbb}}
. The results show that they are smaller than 1-loop contribution by 3 orders and 2 orders of magnitudes respectively, due to small $\epsilon_q$ and an extra loop suppression. In the following section we give an example of a possible UV completion model and discuss the details of the 
experimental constraints on the model.

\begin{table}[htb]
	\begin{tabular}{|c|c|c||c|c||c|c|c|c|}
		\hline
		$m_{\phi}$  [MeV] & $\epsilon_q$ & $\epsilon_\ell$ & ${\rm BR}(K_L\rightarrow\pi^0\phi)$ & $\tau$ [s] & $\tan\beta $
		& $\sin\alpha$ & $ \sin \theta_{1\phi} $ & $\sin \theta_{2\phi} $\\
		\hline
		50 & $1.6\times 10^{-2}$ & $1.22$ &$ 1.7\times10^{-6}$ & $5.1\times 10^{-11}$
		& 100 & $-0.01$& 0.0122 & $1.6\times 10^{-2}$ \\  
		\hline 
		60 & $6.8\times 10^{-3}$ & $0.87$ & $3.2\times 10^{-7}$ & $8.25\times 10^{-11}$
		& 100 & $-0.01$& 0.0087 & $6.8\times 10^{-3}$  \\
		\hline
	\end{tabular}
	\caption{Benchmark points of the effective model which satisfies both the KOTO and $(g-2)_{\mu}$ anomalies. The benchmark with $m_{\phi}=50$ MeV is indicated in Figs. \ref{fig:major-result} \& \ref{fig:major-result-2} by a star. The benchmark with $m_{\phi}=60$ MeV is indicated in Figs. \ref{fig:major-result} \& \ref{fig:major-result-2} by a circle.
	The first three columns are the three free parameters of the effective model, and the fourth and fifth columns are the second set of parameters if one trades the two effective couplings for the Kaon decay branching ratio and $\phi$ lifetime. The last four columns give the corresponding values of $\tan \beta$ and three mixing angles in the UV model.
	}
	\label{tab:eff-model_benchmark}
\end{table}

\section{The UV completion of the model}
\label{sec:UVmodel}

\subsection{type-X 2HDM plus singlet scalar model}
\label{sec:typeXmodel}

In this subsection, we discuss the ultra-violet completion of the effective model and the constraints in the next subsection. 
We shall work on a type-X two Higgs doublet model ~\cite{Barger:2009me, Craig:2012vn, Craig:2013hca}, in which the two Higgs 
doublets couple disjointly to either leptons or quarks. 
We will denote $\Phi_2$ as the scalar doublet that couples to quarks, and $\Phi_1$ as the one that couples to leptons. We will assume
the existence of an additional real singlet scalar $\phi$ mixing with the CP-even neutral components of these Higgs doublets. 
This light scalar $\phi$ is exactly the light degree of freedom in the effective model.

A clear advantage of this framework is that no flavor changing neutral currents associated the the scalar fields appear at tree level.
The appearance of a coupling of one scalar to the lepton fields independent of the one to quark fields allows to solve the $(g-2)_\mu$
anomaly simultaneously with the explanation of the observed KOTO excess. Moreover, as emphasized before, the singlet scalar lifetime will be modified
by the presence of the lepton couplings and will be far shorter than the one that would be obtained by a singlet mixing with only
a SM-like Higgs boson, which allows a solution of the KOTO anomaly in a different mass range than the one obtained in previous works \cite{Egana-Ugrinovic:2019wzj, Dev:2019hho}.

The effective Lagrangian density, describing  the interactions of the scalar doublets  and the fermions is
\begin{align}
 \mathcal{L}_{\rm yuk}  & = - \lambda_u \bar{Q} \tilde{\Phi}_2 u_R - \lambda_d \bar{Q} \Phi_2 d_R 
-  \lambda_e \bar{L} \Phi_1 e_R + h.c. \,,
\end{align}
while the scalar potential can be given as,
\begin{align}
V\left(\Phi_1, \Phi_2, \phi_0 \right) & =  \sum_{i = 1, 2} -\mu_i^2 \Phi_i^\dagger \Phi_i + \lambda_i \left( \Phi_i^\dagger \Phi_i \right)^2  - \mu_\phi^2 \phi^2_0 + \lambda_\phi \phi_0^4 \nonumber \\
& - \left( \rho \phi_0 \Phi_1^\dagger \Phi_2 + H.c.  \right)
+ \lambda_{12} \Phi_1^\dagger \Phi_1 \Phi_2^\dagger \Phi_2 
- \rho_{1\phi} \Phi_1^\dagger \Phi_1 \phi_0  - \rho_{2\phi} \Phi_2^\dagger \Phi_2 \phi_0 \\
& + \dots \nonumber
\label{eq:scalarpotential}
\end{align}
where the first line determines the masses of the three CP-even neutral scalars, the second line determines the three mixing angles between the CP-even neutral scalars and the third line contains all terms irrelevant to our discussion.
The vacuum expectation values (vevs) of the two Higgs doublets are $v_1$ and $v_2$ respectively. We will assume that $\tan \beta = v_2/v_1 \gg 1$, 
implying that the field $\Phi_2$ will have approximately standard interactions with quarks and gauge bosons, and will carry the dominant component of the SM-like Higgs in the alignment limit \cite{Gunion:2002zf,Craig:2013hca, Carena:2013ooa}. We will assume that the non-standard Higgs' have masses around several hundreds of GeV.
 
The CP-even neutral scalars in the two Higgs doublets ($\Phi_{1,2}^0$) will mix with the singlet scalar $\phi_0$ through the mixing matrix
\begin{align}
\begin{pmatrix}
\sqrt{2} \text{Re}\left[\Phi_1^0 \right]        \\
\sqrt{2} \text{Re}\left[\Phi_2^0 \right]  \\
\phi_0
\end{pmatrix}	 \simeq
\begin{pmatrix}
	\cos\alpha          & -\sin \alpha  & \sin \theta_{1\phi}\\
	\sin \alpha        & \cos\alpha & \sin \theta_{2\phi} \\
	- \sin \theta_{1\phi} & -\sin \theta_{2\phi} & 1
\end{pmatrix} \,.	
\begin{pmatrix}
H           \\
h \\
\phi
\end{pmatrix} , 
\end{align}
where we have assumed the mixing angles between the singlet and the CP-even scalars coming from the Higgs doublets
are very small. The mass eigenstates are $h, ~ H , ~ \phi$, where $h \approx \sqrt{2} \text{Re}\left[\Phi_2^0 \right]$ is the SM-like Higgs. 
 In terms of mass eigenstates, the Yukawa lagrangian at leading order in the mixing angles is given by
\begin{align}
\mathcal{L}_{\rm yuk} & \simeq   \left(\sin\alpha \  H +  \cos\alpha \ h + \sin \theta_{2\phi} \ \phi \right)  \sum_{q} \frac{m_q}{v_2} \bar{q} q
+
\left( \cos\alpha \ H - \sin \alpha \ h + \sin \theta_{1\phi} \  \phi \right) \sum_{\ell} \frac{m_\ell}{v_1} \bar{\ell} \ell \nonumber \\
&+ \frac{i}{v} A\left[ m_d \bar {d} \gamma_5 d  \cot \beta -m_u \bar {u}\gamma_5 u \cot \beta  - m_\ell  \bar{\ell} \gamma_5\ell \tan \beta \right] \nonumber \\
& + \frac{\sqrt{2}}{v}m_d H^+\bar{u}_L d_R \cot\beta  -\frac{\sqrt{2}}{v} m_u H^- \bar {d}_L u_R \cot\beta  -\frac{\sqrt{2}}{v}  m_\ell H^+\bar{\nu}_L \ell_R \tan\beta + h.c., 
\end{align}
where $v_1^2 + v_2^2=v^2=(246 \text{ GeV})^2$.\\
For $h$ to be SM-like, its couplings to leptons requires
\begin{align}
\sin\alpha/\cos\beta \approx  - 1 , 
\end{align}
which is the same as the usual Higgs alignment requirement \cite{Gunion:2002zf,  Craig:2013hca, Carena:2013ooa}.

Once we apply this requirement, the Yukawa terms can be simplified as
\begin{align}
\mathcal{L}_{ \rm yuk} 
& \simeq \left( h - \frac{H}{\tan\beta} + \frac{\sin \theta_{2\phi} }{\sin\beta} \phi \right) \sum_{q} \frac{m_q}{v} \bar{q} q
+ \left(  h +  \tan\beta \ H + \frac{ \sin \theta_{1\phi}  }{\cos \beta} \phi \right) \sum_{\ell} \frac{m_\ell}{v} \bar{\ell} \ell 
\end{align}
where we will assume that $\tan \beta \gg 1$. One can read out the relation between UV parameters and the effective model parameters,
\begin{align}
	\epsilon_q \simeq \frac{\sin \theta_{2\phi}}{\sin\beta} , \quad \epsilon_{\ell} \simeq \frac{\sin \theta_{1\phi}}{\cos \beta}.
\end{align}

To calculate the diphoton decay BR for $\phi$, one also needs the coupling to $W$ gauge boson,
\begin{align}
	\epsilon_W & \simeq  \left(  \sin \theta_{1\phi} \cos\beta + \sin \theta_{2\phi} \sin \beta  \right) \\
	& \approx \epsilon_{\ell} \cos^{2} \beta + \epsilon_q \sin^2 \beta \approx \epsilon_q ,
\end{align}
where in the second line, $\tan^2 \beta \gg \epsilon_{\ell}/\epsilon_q$ was assumed. One can see that in this limit, we have simplified to 
$\epsilon_W \approx \epsilon_q $ which is similar to the SSM model case. The charged Higgs can also contribute to the diphoton decay for
$\phi$. However, there are too many parameters in the scalar potential to uniquely determine the relevant coupling. For example,
the mixing angle $\sin \theta_{1\phi}$ can come from $\rho \phi_0 \Phi_1^\dagger \Phi_2$ term instead of $\rho_{1\phi} \phi_0 \Phi_1^\dagger \Phi_1$. As a result, the coupling of $\phi$ to the charged Higgs can be small without changing the phenomenology we are interested in. Therefore we neglect the charged Higgs contribution in our results.

A few comments are in order regarding the calculation of $K_L \to \pi^0 \phi$ in the full model. Recall that this decay is induced by a 1-loop $s-d-\phi$ coupling in the effective model. Similar loop induced decays have been  calculated for $h$ and $H$ in general 2HDM~\mbox{
\cite{Bejar:2000ub, Bejar:2003em}}
. Upon adding the W boson, charged Goldstones and charged Higgs diagrams, the UV cancellation and gauge invariance is explicit. The singlet contribution just proceeds from the small mixing with the $h$ and $H$ Higgs bosons.
Secondly, our calculation in the effective model assumes that the singlet contribution is approximately given by the one coming from the mixing of the singlet with $h$, $\epsilon_q = \epsilon_W$. In the full model, this assumption is violated due to the non-standard Higgs contributions, which are however quite small. For instance  $\epsilon_q + \epsilon_{\ell} /\tan^2\beta \approx \epsilon_W$, implying that the realtion $\epsilon_q = \epsilon_W$ is precise at the $10^{-2}$ level. The contributions coming from the mixing of the singlet with $H$ are suppresed due to the $1/\tan\beta$ suppression of the $H$ coupling to quarks and its suppression to its coupling to the $W$ and Goldstone bosons in the alignment limit, $\cos(\beta-\alpha)\sim 0$. Similarly, the inclusion of the charged Higgs in the diagram is suppressed by three reasons: the $1/\tan\beta$ suppression of the charged Higgs coupling to quarks, the proximity to the alignment limit we are assuming, that suppresses the coupling of the Standard Model Higgs to the charged Higgs and W boson $\cos(\beta-\alpha) \sim 0$, and the heaviness of the charged Higgs. Therefore, we conclude all contributions not contained in our approximate expression are suppressed by at least a $1/\tan\beta$ factor with respect to the ones we considered.  

\subsection{The various constraints}

In the UV model, since $\Phi_{1,2}$ couples to leptons and quarks respectively, there are no FCNC issues at the tree-level.
However, there are constraints from collider, beam dump and astrophysical experiments \cite{Bezrukov:2009yw, Schmidt-Hoberg:2013hba, Clarke:2013aya, Dolan:2014ska, Alekhin:2015byh, Flacke:2016szy}. We will discuss them one by one in the following subsections.

\subsubsection{Collider constraints}
We consider several relevant constraints for the UV model from collider measurements. The first is the modification of the $\tau$ leptonic decay branching ratio, because the  charged Higgs couples to $\tau$ leptons with a large Yukawa coupling. The next is exotic Z decays to lepton final states. The third is the exotic Higgs decay $h \to \phi \phi$ and the last two are B meson decays $B_s \to \mu^+ \mu^-$ and $B\to K^{(*)} e^+ e^-$.   

Firstly, the charged Higgs coupling to leptons is contrained by the $\tau$ leptonic decay~\cite{Krawczyk:2004na, Craig:2013hca}.
Its contribution to the $\tau$ leptonic decay width is given as 
\begin{align}
\Gamma^{H^\pm}_{\rm tree} \left(\tau \to \ell \bar{\nu}_{\ell} \nu_{\tau} \right) =\Gamma_0\left[ \frac{m^2_\tau m^2_\ell \tan^4\beta}{4m^4_{H^\pm}} -2\frac{m_\ell m_\tau \tan^2\beta}{m^2_{H^\pm}}\frac{m_\ell}{m_\tau}\kappa \left( \frac{m^2_\ell}{m^2_\tau}\right) \right],
\end{align}
where $\Gamma_0$ is the SM total decay width of $\tau$ and 
\begin{align}
\kappa(x)=\frac{g(x)}{f(x)}, \quad & g(x)=1+9x-9x^2-x^3+6x(1+x)\ln{x}, 
\nonumber\\
& f(x)=1-8x+8x^3-x^4-12x^2\ln{x} .
\end{align}
It is clear that if the combination $m_\tau m_\ell \tan^2\beta \ll m^2_{H^\pm}$, then the deviation
from the SM decay width is small. The SM branching ratio for the tau decay to a muon plus a neutrino pair is measured to be
${\rm BR}\left(\tau \to \mu \bar{\nu}_\mu \nu_\tau \right) = 17.39 \pm 0.04 \%$ \cite{Tanabashi:2018oca}. 
Therefore, within $1\sigma$, this measurement constrains the combination $\tan^2\beta/m^2_{H^\pm} < 0.89 ~{\rm GeV^{-2}}$.
For our benchmark points we have taken $\tan \beta = 100$. Thus, the constraint from tau decays requires that $m_{H^\pm } \gtrsim 100$ GeV,
which is easily  satisfied. On the other hand, the contribution of $H^\pm$ for $\tau \to e \bar{\nu}_e \nu_\tau$ is suppressed 
by the electron mass, thus the corresponding constraint is much weaker.

Secondly, since the new Higgs couplings to leptons are enhanced by large $\tan \beta$, one might worry that the 
branching ratio $Z \to 4\ell$ might be changed due to new Higgs mediation. The measurement on Z decay branching ratio $Z \to 4\ell$
is $\left(4.58 \pm 0.26 \right) \times 10^{-6}$ \cite{Rainbolt:2018axw}, where $\ell$ means $e, ~ \mu$.
However, in their study, an invariant mass cut larger than 4 GeV is applied  for all opposite-sign, same-flavor lepton pairs.
Therefore, the exotic decay $Z \to \mu^+ \mu^- \phi$ does not set a significant constraint.  

Thirdly, the exotic Higgs can decay to a pair of $\phi$ particles, $h \to \phi \phi$.  Since $\phi$ decays to $e^+ e^-$ and has a mass around $50 $ MeV, the electron and positron will be highly collimated and can not be separated by the LHC. Such an exotic event might be identified as $h \to e^+ e^-$ and could be constrained by exotic Higgs decays to two lepton jets measurements \cite{Aad:2015sms}. The ATLAS collaboration performed a search for prompt lepton jets at 8 TeV and the model they consider is $h \to A' A' + X$ where the two $A' $ decay to lepton jets subsequently. For $m_{A'} = 0.4$ GeV, the constraint with prompt $A'$ decay is ${\rm BR}(h \to A' A' + X) \lesssim 0.3 \%$ at $95\%$ C.L.. Since $\phi$ decays mostly to an electron pair in our model, it falls into the electron lepton jet which is less stringent. It is reasonable to say that the constraint for ${\rm BR}( h \to \phi \phi) \lesssim 1 \%$. Again, in the scalar potential, there are many free parameters and the relevant terms in the second line of Eq.~(\ref{eq:scalarpotential}) will control the exact value of the  $\phi-\phi-h$ coupling.  Therefore, this constraint may be avoided 
while keeping the three mixing angles required to explain the KOTO and $(g-2)_\mu$ anomalies.

Next, $\phi$ can mediate $B_s \to \mu^+ \mu^- $ decays as it couples, in particular to bottom and strange quarks, and muons. Recently, this process has been measured by CMS and LHCb \cite{Aaij:2013aka, Chatrchyan:2013bka,CMS:2014xfa,Aaij:2017vad,Aaboud:2018mst} and found to be in agreement with the SM predictions. The constraints for a single-mixing coupling with strength proportional to $m_f/v$ are calculated for a pseudoscalar in \cite{Dolan:2014ska} and for a scalar in \cite{Alekhin:2015byh, Flacke:2016szy,Li:2014fea,Arnan:2017lxi}, which conservatively require that the new contribution can not exceed the SM branching ratio. Relevant to our scenario, for $m_\phi < 1$ GeV, the limit on $\epsilon_f$ is $\lesssim 1$. Since the process is proportional to $\epsilon_f^4$ \cite{Altmannshofer:2011gn}, in our model the corresponding factor is $\epsilon_q^2 \epsilon^2_\ell$. Fitting the $(g-2)_\mu$ anomaly implies $\epsilon_\ell \sim 1$ for the light $\phi$, but this constraint becomes quite weak since $\epsilon_q$ is about $10^{-3}$--$10^{-2}$. 
One might worry about the heavy Higgs $H$ which carries a $\tan\beta$ enhanced lepton coupling. However, its quark coupling is suppressed by the same $\tan\beta$ factor and thus the multiplication of quark and lepton couplings gives, apart from a sign, the same as that for the SM-like Higgs $h$. Given the fact that the $H$ is heavier than $h$ and the SM contribution is dominated by Z-penguin and box diagram, it is safe to neglect the heavy Higgs boson contribution to this process. Similarly the couplings of fermions to the charged Higgs are suppressed by $1/\tan\beta$ together with the charged Higgs mass, thus we neglect this contribution as well.

Lastly, $\phi$ can result in the B meson decay $B \to K^{(*)} \phi$, where $\phi$ subsequently decays to $e^+ e^-$. The branching ratio measurement of ${\rm BR} (B \to K^{(*)} e^+e^-)$ can generally constrain the 2HDM plus singlet scalar model, see \mbox{
\cite{Datta:2019bzu}}
. For our benchmarks, one can calculate the branching ratio of $B \to K^{(*)} \phi$ to be $\sim 10^{-4}$, following the calculations in \mbox{
\cite{Batell:2009jf}}
.
The LHCb measurement of the SM branching ratio ${\rm BR} (B^0 \to K^{(*0)} e^+e^-)$ is about $3.1^{+0.94}_{-0.88}\times 10^{-7}$ in the dilepton mass range from 30 MeV to 1000 MeV \mbox{
\cite{Aaij:2013hha}}
. However, the reconstruction of $B^0$ needs $K^{*0}$ and $e^+e^-$ to form a good quality vertex, where the LHCb vertex resolution is about $\sigma_v^T = \mathcal{O}(10) ~\mu {\rm m}$ in transverse plane and $\sigma_v^L = \mathcal{O}(100) ~\mu {\rm m}$ in longitudinal direction, highly depending on the number of tracks of the vertex \mbox{
\cite{LHCbVELOGroup:2014uea}}
. Our benchmarks have a lifetime of $c\tau \sim 1.5$ cm, therefore, the probability for $\phi$ to decay within the vertex resolution $L$ is $1 - e^{- m_\phi L /(p_\phi \tau_\phi)} \sim m_\phi L / (p_\phi \tau_\phi)$. The average momentum of the B meson is about 80 GeV \mbox{
\cite{Altarelli:2008xy}}
, thus the decay probability is $\sim 4 \times 10^{-4}$, conservatively taking $L \sim 5$ mm \footnote{Although the vertex resolution of LHCb is about $\mathcal{O}(100) ~\mu {\rm m} $, the references \mbox{
\cite{Schmidt-Hoberg:2013hba, Dolan:2014ska, Winkler:2018qyg} }
choose $L \sim 5$ mm as the criteria of prompt decay.} and $p_\phi \sim 40$~GeV. Thus, this decay is too small to contribute to the LHCb measurement. Moreover, in the 3-body decay $B^0 \to K^{(*0)} e^+e^-$ in SM, the angle between $e^+e^-$ can be quite large. However, in our case, $e^+e^-$ is very collimated giving a signature in the detector similar to the SM background $B^0 \to K^{*0} \gamma$, when the photon later converts into an $e^+ e^-$ pair. As a result, we conclude our benchmark is safe from the prompt search of 
${\rm BR} (B^0 \to K^{(*0)} e^+e^-)$ due to large lifetime and small scalar mass. 
It is worth mentioning that the BaBar collaboration has considered $B \to X_s\phi$ decays where $X_s$ is the strange hadronic system, with subsequent displaced decays $\phi \to e^+ e^-, ~ \mu^+ \mu^-$ and several other possible channels \mbox{
\cite{Lees:2015rxq}}
, however $m_{ee} > 0.4$ GeV is required.

\subsubsection{Beam dump experiments}

In this section, we consider various beam dump experiments. In this type of experiment, displaced decays of new particles is critical,
otherwise they will be swamped by the SM background. Therefore, the lifetime is a very important factor in the signal analysis as it appears in the exponential form of the decay probability. Some of the beam dump experiments look for new particle decays to visible final states, e.g. electron pairs, while some look for invisible decays. In our scenario, the light scalar $\phi$ will dominantly decay to visible final states $e^+ e^-$ and subdominantly to $\gamma \gamma$. With a finite lifetime, it can always be subject to constraints by either kind of beam dump experiments, depending on decay probability within and without the detector. We will go through the various experiments in the following discussion.  \\

\noindent \textbf{E949 and NA62:} 

The E949 collaboration \cite{Anisimovsky:2004hr, Artamonov:2009sz} and NA62 collaboration \cite{NA62-2019} have measured the process $K^+ \to \pi^+ \bar{\nu} \nu$, which could be mimicked by the $K^+ \to \pi^+ \phi$ decay. 

The recent NA62 result combines 2016 and 2017 data, which sets ${\rm BR} \left(K^+ \to \pi^+ \bar{\nu} \nu \right) < 1.85 \times 10^{-10}$ at $90 \% $ C.L. If the process is the two body decay $K^+ \to \pi^+ \pi^0$, they can also constrain ${\rm BR}(\pi^0 \to {\rm invisible } ) < 4.4 \times 10^{-9}$ at $90 \% $ C.L.
Since the $\phi$ has finite lifetime, there is probability for it to decay outside the detector. Therefore, one can use Eq.~(\ref{eq:BR_KOTO}) to describe the constraint for NA62, with the substitution of detector size $L = 150$ m. Since NA62 used the flying Kaon with energy of 75 GeV, one can calculate the energy of $\phi$, assuming an isotropic decay of $K^+$ and further requiring that the $\pi^+$ energy falls into their signal box $15$--$35$ GeV.
Neglecting the signal efficiency difference between 2-body and 3-body decay of $K^+$, we can arrive at a constraint for NA62 on the parameter space, which is the purple shaded region in Fig.  \ref{fig:major-result} and \ref{fig:major-result-2}. 

For the result from the E949 collaboration \cite{Artamonov:2009sz}, the possibility of $K^+ \to \pi^+ \phi$ with $\phi$ being long-lived has been explicitly explored. The constraints on ${\rm BR}(K^+ \to \pi^+ \phi)$ have been given as a function of $m_\phi$ and its lifetime. We translate the limit into our signal model and the excluded parameter space is shown as blue shaded region in Fig. \ref{fig:major-result}.
We comment on the constraint from the invisible decay for B meson $B \to K + {\rm inv}$. It can exclude $\epsilon_q \lesssim 10^{-2}$ in SSM model \cite{Clarke:2013aya, Alekhin:2015byh, Flacke:2016szy}, however it can not be applied in our case because $\phi$ is too short-lived. 
\\

\noindent \textbf{CHARM: }

The CHARM experiment measures the displaced decay of neutral particles into $\gamma \gamma , ~ e^+e^-$ and $\mu^+\mu^-$. Since our signal can result from $\phi$ being produced from the decays $K_L \to \pi^0 \phi$ and $K^+ \to \pi^+ \phi$, the CHARM experiment is
relevant for long-lived $\phi$. Following \cite{Dolan:2014ska}, we have the number of events of this exotic decay for CHARM detector to be
\begin{align}
N_{\rm det} \approx  N_\phi \left( e^{- \frac{480 m}{ \gamma_\phi \beta_\phi c\tau_\phi}} - e^{- \frac{480 +35 m}{ \gamma_\phi \beta_\phi c\tau_\phi}}  \right)
\sum_{X = e,\mu,\gamma} {\rm BR}\left(\phi \to X X \right).
\label{eq:charm}
\end{align}
In CHARM experiment, the energy of $\phi$ is about 10 GeV thus one can calculate the velocity $\beta_\phi$ and the boost factor $\gamma_\phi$.
The number of $\phi$ produced in the Kaon decay is $N_\phi = 2.9 *10^{17} \sigma_\phi/\sigma_{\pi_0}$  and $\sigma_\phi$ is production cross-section \cite{Bezrukov:2009yw} 
\begin{align}
\sigma_\phi \approx \sigma_{pp} M_{pp} \chi_s \left( 0.5 ~ {\rm BR} \left(K^{+ } \to \pi^+ \phi \right)  + 0.25 ~ {\rm BR} \left(K_{L} \to \pi^0 \phi \right)\right),
\end{align}
where $\sigma_{pp} $ is the total proton cross-section, $M_{pp}$ is the total hadron multiplicity and $\chi_s =1/7$ is the relative parts going into strange flavour \cite{Andersson:1983ia}. For the pion, there is the relation $\sigma_{\pi_0} \simeq \sigma_{pp} M_{pp} / 3$. 
In \cite{Dolan:2014ska}, the authors calculated the CHARM limit for $\phi$ as a pseudoscalar. In \cite{Clarke:2013aya, Alekhin:2015byh, Flacke:2016szy}, the CP-even scalar case is calculated, and the effect of the parity of the scalar is subdominant. 
Since the CHARM experiment has observed zero event for such exotic decay, one can set $90\%$ confidential level bound by 
requiring $N_{\rm det} < 2.3$. The CHARM experiment excludes the parameter space of our model in the yellow shaded region in Fig. \ref{fig:major-result} and \ref{fig:major-result-2}.
\\

\noindent \textbf{$K_{\mu 2}$ experiment:}

Due to $K^+ \to \pi^+ \phi$ process, a stopping $K^+$ decay to $\pi^+$ in two body final state is relevant as a by product of the $K_{\mu 2}$ experiment~\cite{Yamazaki:1984vg}. Because $K^+$ is stopped in the above 2-body decay, the momentum of $\pi^+$ is mono-chromatic with a dependence only on the $\phi$ mass. Since this energy peak has not been observed, one can set a constraint on ${\rm BR}\left( K^+ \to \pi^+ \phi \right)$, regardless of  the $\phi$ decay products~\cite{Yamazaki:1984vg, Dolan:2014ska, Alekhin:2015byh}. We apply this constraint to our model and translate the constraint on ${\rm BR}\left( K_L \to \pi^0 \phi \right)$ to the model parameter $\epsilon_q$.  The excluded parameter space is shown as the cyan shaded region in Fig.  \ref{fig:major-result} and \ref{fig:major-result-2}.
\\

\noindent \textbf{KTeV/E799 ($ e^+ e^-$):}

Given the $\phi$ mass we are interested in, $\phi$ will dominantly decay to $e^+ e^-$. The searches for $K_L \to \pi^0 e^+ e^-$ from  KTeV/E799 \cite{AlaviHarati:2003mr} has set limits ${\rm BR} \left( K_L \to \pi^0 e^+ e^-\right) < 2.8\times 10^{-10} $ at $90\%$ C.L. , which is relevant for the process $K_L \to \pi^0 \phi$. 
Another search $K^+ \to \pi^+ e^+ e^-$ from NA48/2 \cite{Batley:2009aa} has measured ${\rm BR} \left( K^+ \to \pi^+ e^+ e^-\right) = (3.11 \pm 0.12)\times 10^{-7} $, which is also relevant to our signal, but is much less stringent.
In these type of searches, an invariant mass cut of $m_{e^+e^-} > 140$ MeV is always applied to suppress the dominant background $K \to \pi \pi^0_D$, where $\pi^0_D$ is the pion Dalitz decay $\pi^0 \to \gamma e^+ e^-$. Therefore it is not applicable for light $\phi$ with $m_\phi < 140$ MeV. Furthermore, the leptons $e^+e^-$ are required to have a common vertex with the Kaon decay vertex. Thus, if $\phi$ is long-lived the constraint will vanish. Following~\cite{Dolan:2014ska}, the vertex resolution of KTeV/E799 is taken to be 4 mm, which corresponds to $1.3 \times 10^{-11}$ seconds. The probability of $\phi \to e^+ e^-$ being prompt is 
\begin{align}
P_{\rm prompt} = 1 - e^{- \frac{L_{\min}}{\tau_{\phi}}  \frac{m_{\phi}}{p_{\phi}}},
\end{align}
where $L_{\min}$ is the vertex resolution, $p_\phi$ is the momentum of $\phi$ in the laboratory frame in the decay. For the KTeV/E799 experiment, the Kaon has an energy range of $20.3$--$216$ GeV \cite{AlaviHarati:2003mr}, which determines the energy of $\phi$ when assuming isotropic decay in the Kaon center of mass frame. Further, assuming the Kaon energy has a flat distribution, one can calculate the probability $P_{\rm prompt}$.
We set the KTeV/E799 constraint in the orange shaded region in Fig. \ref{fig:major-result}. 
\\

\noindent \textbf{KTeV ($\gamma\gamma$):}

The KTeV collaboration has measured the process $K_L \to \pi^0 \gamma \gamma$ \cite{Abouzaid:2008xm},  and determined the branching ratio to be ${\rm BR}\left( K_L \to \pi^0 \gamma \gamma \right) = \left( 1.29 \pm 0.03 \pm 0.05 \right) \times 10^{-6}$. Since the four photons are measured by the CsI calorimeter, the information is not sufficient to reconstruct the decay vertex as it needs to assume that the invariant mass of four photons equals the Kaon mass. Therefore, it is not sensitive to whether the $\phi \to \gamma \gamma$ decay is displaced or prompt in the signal process $K_L \to \pi^0 \phi$. The efficiency difference between 3-body decay and 2-body decay $K_L \to \pi^0 \phi$ with $\phi \to \gamma \gamma$ is not given in \cite{Abouzaid:2008xm}. As a result, we can conservatively assume that the new physics contribution should be as small as ${\rm BR}\left( K_L \to \pi^0 \phi \right) {\rm BR}\left( \phi \to \gamma \gamma \right) \lesssim 10^{-6}$, see \cite{Kitahara:2019lws}.
The constraint for KTeV shown in the brown shaded region in Fig. \ref{fig:major-result} and \ref{fig:major-result-2}. 
\\
 
\noindent \textbf{Orsay:}

Orsay is an electron beam dump experiment which is sensitive to a light scalar decaying to electrons. The process is electron bremsstrahlung $e N \to e N \phi$ where $\phi $ subsequently decays to $e^+ e^-$. In \cite{Davier:1989wz}, a search for light Higgs bosons was performed under the assumption that the light Higgs couplings exclusively to electrons. This constraint, at the $90 \%$ C.L., can be directly applied to our model, where the only difference is a tiny 
$\rm BR(\phi \to \gamma\gamma)$ which can be neglected for $m_{\phi} < 60$ MeV. We fit to the $90 \%$ C.L. constraint set in \cite{Davier:1989wz} and project this limit to the $95 \%$ C.L. The results are shown in Fig. \ref{fig:major-result} and Fig. \ref{fig:major-result-2} in the green shaded region and the fitting procedure is described in detail in the Appendix.
\\

\noindent \textbf{E137:}

Additional constraints are also relevant from other electron beam dump experiments at SLAC, e.g. E137 \cite{Bjorken:1988as} and E141 \cite{Riordan:1987aw}. In \cite{Bjorken:2009mm, Andreas:2012mt, Bauer:2018onh}, these experiments were used to constrain the parameter space for dark photons. In particular, the E137 experimental setup provides an accurate upper bound on the dark-photon kinetic mixing. Therefore, we estimate the E137 limit by translating the bound on kinetic mixing to the scalar lifetime by equating the corresponding lifetime of the dark photon, $\tau_{A'} = \tau_{\phi}$. This limit is shown in Fig. \ref{fig:major-result} and Fig. \ref{fig:major-result-2} in light gray shaded region.  Indeed, suppose we can split the signal event into $N \approx \sigma P(\tau)$, where $\sigma$ is the production cross-section, and $P(\tau)$ is the probability to decay within the right volume. Assuming that $\tau_{A'} = \tau_{\phi}$, the probabilities for decay will be equivalent $P_{A'} = P_{\phi}$. This would imply that $\epsilon e > \epsilon_{ \ell} m_e/v$ since
\begin{equation}
\Gamma(A' \to e^+e^-) \approx \frac{(\epsilon e)^2}{12\pi} m_{A'}.
\end{equation}
 Comparing to Eq. \ref{eq:diphoton} this gives $\sigma_{A'} > \sigma_{\phi}$. Though, the difference in cross sections occurs even in the case when the couplings are equivalent due to the collinear enhancement of the vector boson production rate. Thus, the number of events  which decay within the detector in the case of scalar particles is strictly smaller than that for dark photons. Hence, while translating the bounds on the dark photon mass for a given lifetime to the scalar case, we are imposing a conservative bound.
 
 We have also compared our results to the limit presented in \cite{Batell:2016ove}, where the limit is obtained by relating the dark photon kinetic mixing and Higgs-like couplings, $\epsilon e = \epsilon_{ \ell} m_e/v$. We find that our estimate is slightly weaker than the one obtained by this method. However, we argue that relating the dark photon and scalar lifetimes provides a more accurate bound. Indeed,
 assuming that $\epsilon e = \epsilon_{ \ell} m_e/v$ one obtains $\tau_{A'} > \tau_{\phi}$, or  $P_{A'} > P_{\phi}$. In addition, the larger cross section associated to the dark photon production lead to a further enhancement of their event rate, implying that their  bound is too restrictive when applied to the scalar case.  

\subsubsection{Astrophysical constraints}

Astrophysical constraints can set relevant bounds on models with long-lived scalars. For instance,
supernova can loose energy by emitting the light scalar outside the neutrino sphere $R_\nu =40$ km \cite{Chang:2016ntp}. If the scalar decays inside or is absorbed inside $R_\nu$, the supernova neutrino flux is not affected. The supernova can not stand the instantaneous luminosity of exotic particle emission, which exceeds the neutrino luminosity when the core reaches its
peak density $\rho_c \sim 3\times 10^{14} ~{\rm g/cm^3}$ and temperature $T_c \sim 30 ~{\rm MeV}$. Otherwise,
the duration of the neutrino burst will be shortened by half and the energy spectrum will be incorrect \cite{Raffelt:1996wa}.
The typical core radius $R_c$ is about 10 km. Since in our interested region, the scalar $\phi$ has mass around $\mathcal{O}(30)$ MeV and lifetime around $10^{-11}$--$10^{-10}$ seconds, the decay length is about $0.3$--$3$ cm, and hence much smaller than the size of the core. 
Naively, such decay length is too small to result in a sizable energy leak due to $\phi$ emission.

Before discarding this constraint, one should remember that the progenitor star is an electron rich environment,  and hence it is necessary to consider the Pauli blocking effect which suppresses $\phi$ decaying to electron pairs. Inside the core, the chemical potential for the electron is $\mu_e \simeq 100$ MeV \cite{Rrapaj:2015wgs}. The Pauli blocking factor is 
\begin{align}
 \left(e^{\frac{\mu_e - E_e}{T}}+1\right)^{-1} ,
\end{align} 
where we can conservatively replace the electron energy $E_e$ as $m_{\phi}/2$.   
Due to the high $T_c$ in the core, the Pauli blocking factor is about 0.05 which is too small to make $\phi$ decay outside of the core. Outside of the core region, the Pauli blocking factor can also be calculated. Adopting fiducial model parameters for the progenitor star \cite{Raffelt:1996wa, Chang:2016ntp}, the density and temperature can be modeled as $\rho(r) = \rho_c (r/R_c)^{-5}$ and $T(r) = T (r/R_c)^{-5/3}$. Applying a uniform proton factor of $0.3$, one can obtain the number density of electron $n_e$, which equals the number density of proton $n_p$. The temperature within the neutrino sphere $R_\nu$ is higher than 3 MeV \cite{Chang:2016ntp}, thus the electrons are  relativistic. In the limit of $m_e = 0$, one can have a relation $n_e(T, \mu_e) = (3 \pi^2)^{-1} \mu_e (\mu_e^2 + \pi^2 T^2)$ \cite{Braaten:1993jw}. Solving for the electron chemical potential, we find that the Pauli blocking factor is about $3\times 10^{-4}$ for $R_c<r<R_\nu$. Taking account the Pauli blocking, the typical decay length for $\phi$ becomes $10$--$100$ meters, which is still much smaller than the size of the neutrino sphere $R_\nu$.
Therefore, unlike the case of the SSM model in~\cite{Dev:2019hho}, in our scenario the supernova constraint is not relevant due to the short lifetime of~$\phi$.

Besides the supernova, the light scalar $\phi$ can also affect the big bang nucleosynthesis (BBN) if it decays to SM particles fairly late. In our scenario, $\phi$ couples to leptons with similar a Yukawa coupling as the  Higgs. Thus, its lifetime is as small as $\sim 0.1$ nanosecond, which is much smaller than lifetimes larger than $\mathcal{O}(1 \ {\rm sec})$ that are constrained by these considerations. As a result, our model will not affect either BBN or the cosmic microwave background.

\section{Conclusions}
\label{sec:conclusion}
Although the SM accurately describes all experimental data, it is expected to be only an effective
field theory. The presence of extra scalar degrees of freedom, beyond the standard Higgs doublet, is a natural
feature that is present in many extensions of the SM.  Motivated by experimental data, we have
explored the possible extension of the scalar sector by an additional Higgs doublet and a singlet. The
singlet plays an essential role in our phenomenological analysis and it couples
to quarks, leptons and gauge bosons via the mixing with the Higgs doublets, therefore avoiding
tree level flavor changing neutral interactions.   

Our goal has been to explain at the
same time the observed anomalous magnetic moment of the muon and the excess of events observed at 
the KOTO experiment.  We have shown that for this to happen the singlet should have a mass between about 40 and 70 MeV,
and with a coupling to leptons and to quarks that are fixed by the $(g-2)_\mu$ anomaly and the KOTO excess,
respectively. While the couplings to leptons should be approximately of the same order as the Yukawa coupling
to the SM Higgs boson, the coupling to quarks should be more than two orders smaller than the SM Yukawa coupling. 

Under these conditions, the singlet lifetime is controlled by its decay into electrons, which is by far the dominant
decay mode of this scalar. Actually, the singlet decays significantly more promptly than in the previously explored 
case in which it only mixes with the SM Higgs.  The lifetime that is obtained is of the same 
order as the one needed to invalidate the constraints coming from the charged Kaon decay into charged pions 
and neutrinos, avoiding therefore the Grossman-Nir bound on the analogous neutral Kaon decays. This happens 
for a relatively large range of masses, which is a very attractive feature of our scenario.  

The requirement of obtaining different values of the coupling to quarks and leptons with respect to the SM Yukawa
couplings is obtained by assuming that the 2HDM is of type-X, 
in which one Higgs doublet couples only to quarks and
the other doublet couples to leptons.  We assume the system to be close to the alignment limit, implying
the presence of a scalar with similar couplings to quark, leptons and gauge bosons as the SM Higgs.  We also
assume that $\tan\beta$ is large, of the order of 100, with the doublet that couples to leptons acquiring a small
vacuum expectation value.  This implies that the SM Higgs doublet will be mostly associated with the one that
couples to quarks. This also implies that although the singlet has small mixings with the neutral CP-even
components of both Higgs doublets, its coupling to leptons will be similar to the SM Yukawa due to the large 
$\tan\beta$  enhancement of the lepton coupling to the non-standard Higgs doublet. 

There are two features that do not have a natural explanation in our model, but depend strongly on details of the model
that are not associated with the phenomenological properties discussed in this article. One is the possible
decays of the SM Higgs into a pair of singlets. In order to preserve the agreement with Higgs precision 
measurements, the branching ratio of this decay should be smaller than $10^{-2}$.  The second feature 
is associated with the stability of the singlet mass. In our model the singlet mixing angle to the non-standard 
neutral CP-even Higgs is of the order of $10^{-2}$.  This mixing may be smaller,  but at the cost of increasing the coupling
of the non-standard neutral Higgs bosons to tau leptons and inducing a Landau pole on this coupling at too low scales. 
Although small, this mixing induces corrections to the singlet mass that are much larger than its predicted value. 

Beyond these theoretical issues, the model presented here leads to an explanation of both the 
observed value of $(g-2)_\mu$ and of the KOTO excess, while avoiding the severe proton beam dump experiments and astrophysical constraints.
Due to the relatively large coupling of the new scalar to leptons, some of the strongest constraints on our model come from electron beam dump experiments.
These experiments lead to somewhat weaker bounds for values of $(g-2)_\mu$ below the current measured value, while values of  $(g-2)_\mu$ more than
two standard deviations above the current measured value are firmly ruled out. Therefore, the expected measurement of the muon anomalous 
magnetic moment at the g-2 experiment at Fermilab will further test this model. 
Moreover, the KOTO experiment is expected to update its measurement within the next few years. Therefore, this
model will be tested in a definitive way by the KOTO and g-2 experiments in the near future.

\section*{Acknowledgments}
We would like to thank Alakabha Datta, Elina Fuchs, Samuel D. McDermott, Michael Schmitt, Jessica L. Rainbolt,  Lian-tao Wang and Wen-bin Qian for very useful discussions and communication. 
Work at University of Chicago is supported in part by U.S. Department of Energy grant number DE-FG02-13ER41958. Work at ANL is supported in part by the U.S. Department of Energy under Contract No. DE-AC02-06CH11357. JL acknowledges support by Oehme Fellowship. NM is supported by the U.S. Department of Energy, Office of Science, Office of Work- force Development for Teachers and Scientists, Office of Science Graduate Student Research (SCGSR) program. The SCGSR program is administered by the Oak Ridge Institute for Science and Education (ORISE) for the DOE. ORISE is managed by ORAU under contract number de-sc0014664.

\appendix
\section{Numerical fit of Orsay result}
\label{sec:orsayfit}

\begin{figure}[htb]
 	\includegraphics[width= 0.6 \textwidth]{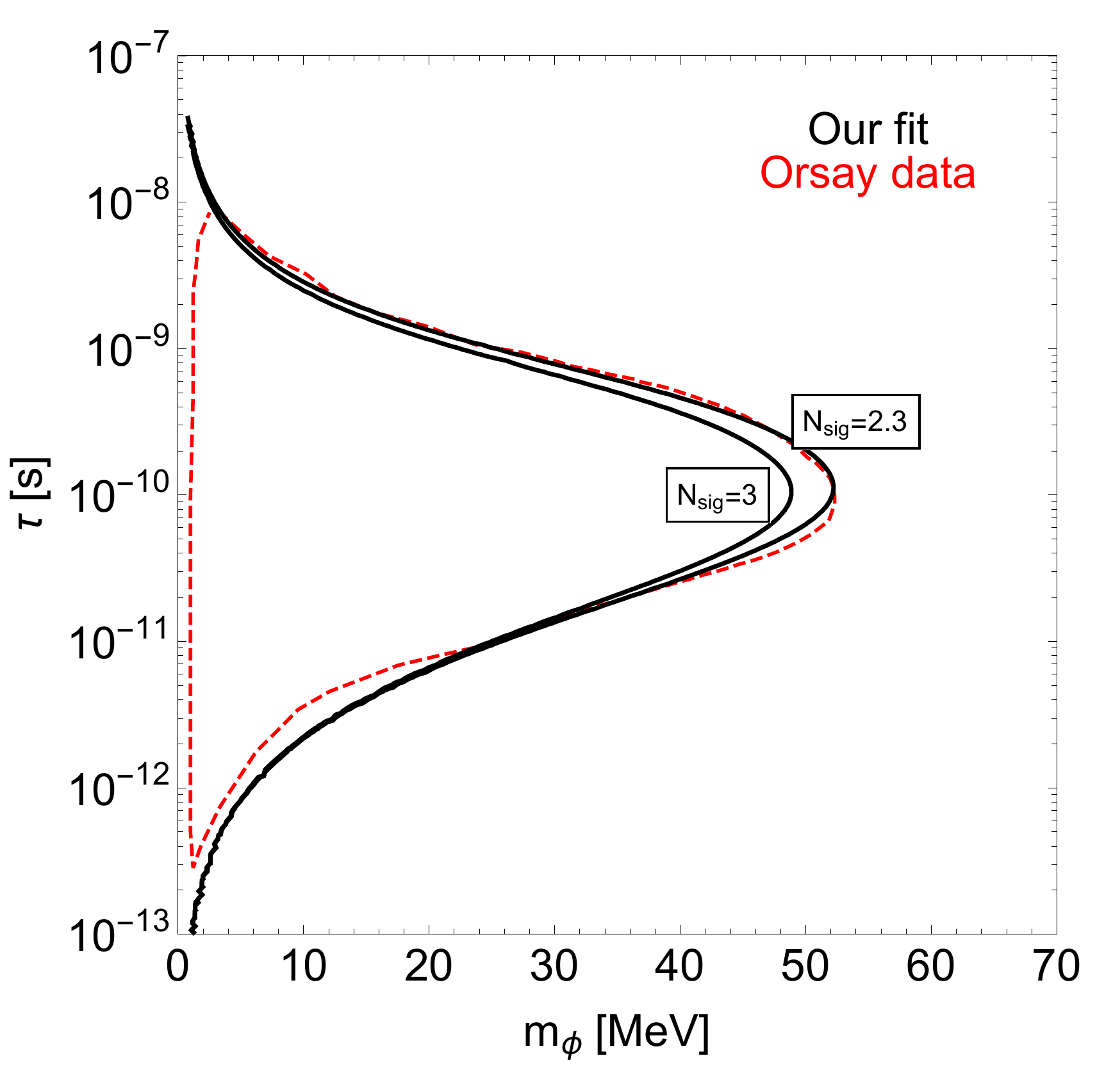}
 	\caption{The numerical fit to Orsay constraint. }
 	\label{fig:fit}
 \end{figure}

In this appendix, we provide the details of our estimation for the $95 \% $ C.L. Orsay constraint \cite{Davier:1989wz}. The signal number for Orsay should have a form close to
\begin{align}
N_{\rm sig} \approx \frac{c_0}{\tau_{\phi}  m_\phi^3} \left(  e^{-a_1 \frac{L_{\rm sh}}{c \tau_\phi \beta_\phi } \frac{m_\phi}{E_\phi}}
-  e^{-a_2 \frac{L_{\rm sh} + L_{\rm dec}  }{c \tau_\phi \beta_\phi } \frac{m_\phi}{E_\phi}} \right)  ,
\label{eq:fitOrsay}
\end{align}
where  the term $(\tau_{\phi}  m_\phi^3)^{-1}$ comes from the total signal production, and the term in parentheses is the decay probability for Orsay, and $L_{\rm sh} = 1$ m, $L_{\rm dec} = 2$ m \cite{Bauer:2018onh}. The energy $E_\phi$ should be within 0.75 GeV and 1.6 GeV, where the upper bound is the electron beam energy and lower bound is the experimental cut \cite{Davier:1989wz}. We reserve $a_1$ and $a_2$ to be $\mathcal{O}(1)$ factors to compensate the electron beam energy attenuation, geometric setup of experiment and signal efficiency of the experiment. $c_0$ is an overall factor which fits to the $90 \% $ C.L. bound ($N_{\rm sig} = 2.3$) given in Fig. 4 of \cite{Davier:1989wz}. In Fig. \ref{fig:fit} we show the fitting result using Eq. \ref{eq:fitOrsay}, and our projection to the $95 \% $ C.L. ($N_{\rm sig} = 3$).

\bibliography{ref}

\bibliographystyle{JHEP}

\end{document}